\documentclass[aps,twocolumn,prb,showpacs,amsmath,amssymb,citeautoscript,nobibnotes]{revtex4}

\usepackage{graphicx}

\newcommand{\lav}{\langle\,}
\newcommand{\rav}{\,\rangle}
\newcommand{\rve}{\boldsymbol r}
\newcommand{\kve}{\boldsymbol k}

\newcommand{\bem}{\begin{em}}
\newcommand{\eem}{\end{em}}
\newcommand{\bq}{\begin{eqnarray}}
\newcommand{\eq}{\end{eqnarray}}
\newcommand{\bqn}{\begin{eqnarray*}}
\newcommand{\eqn}{\end{eqnarray*}}

\newcommand{\beq}{\begin{equation}}
\newcommand{\eeq}{\end{equation}}
\newcommand{\NiY}{Ni$_{33}$Y$_{67}\,$}

\newcommand{\etal}{\bem et~al.\eem}    

\begin{document}
  
\title{Understanding fragility in supercooled Lennard-Jones mixtures.\\ I. Locally preferred structures}
\author{D. Coslovich}
\email{coslo@.ts.infn.it}
\author{G. Pastore}
\email{pastore@.ts.infn.it}
\affiliation{Dipartimento di Fisica Teorica, Universit{\`a} di Trieste
  -- Strada Costiera 11, 34100 Trieste, Italy} 
\affiliation{CNR-INFM Democritos National Simulation Center -- 
  Via Beirut 2-4, 34014 Trieste, Italy} 
\date{\today}
\begin{abstract}
We reveal the existence of systematic variations of isobaric fragility
in different supercooled Lennard-Jones binary mixtures by performing
molecular dynamics simulations. The connection between fragility and
local structures in the bulk is analyzed by means of a Voronoi
construction. We find that clusters of particles belonging to locally
preferred structures form slow, long-lived domains, whose spatial
extension increases by decreasing temperature. As a general rule, a
more rapid growth, upon supercooling, of such domains is associated to
a more pronounced super-Arrhenius behavior, hence to a larger
fragility. 
\end{abstract}

\maketitle

\section{Introduction}

What is the origin of super-Arrhenius behavior of transport
coefficients and relaxation times in supercooled liquids?
This is one of the long-standing, open questions regarding the physics
of the glass-transition. 
Angell~\cite{angell88} introduced the notion of fragility to quantify 
the degree of super-Arrhenius behavior:
generally speaking, the steeper the increase of
relaxation times by decreasing temperature, the more
fragile the glass-former. While relaxation times increase by
several orders of magnitude on approaching the glass-transition, only
mild variations in the average liquid structure are observed.
Nevertheless, a subtle, deep link between microstructural order and
dynamics is believed to exist~\cite{ediger00,widmercooper05,shintani06}. 
In particular, the emergence, upon 
supercooling, of slow domains, characterized by well-defined locally preferred
structures~\cite{frank52}, has been recently identified as a
possible origin of super-Arrhenius behavior, both in numerical
simulations~\cite{dzugutov02,doye03} and theoretical
approaches~\cite{tarjus05,tanaka05b,tanaka05c}. 
Despite some efforts in this direction, the microscopic foundations of
such a connection have been explored only to a limited extent.
In fact, the relation between fragility and local order has been 
explicitly investigated only for monoatomic systems~\cite{dzugutov02,doye03,shintani06}, and operational
schemes for a direct determination of locally
preferred structures have not been tested yet for model supercooled
liquids~\cite{mossa03,mossa06}. 

Understanding how fragility changes in different systems appears to be
an even harder task. Experimentally, some interesting correlations have been
proposed~\cite{ito99,ngai00,scopigno03,novikov05,tanaka05a,niss07}, but 
a sharp interpretation of these results is hindered by the
complexity and the varying nature of intermolecular interactions.
Numerical simulations, allowing fine-tuning of interaction parameters 
for a wide choice of potentials, are optimal tools for studying how the details
of the interaction affect the behavior of supercooled liquids. 
Following this approach, some authors   
have recently investigated the connection between the features of the 
interaction potential and
fragility~\cite{demichele04,bordat04,molinero06,sun06}.
For instance, it has been shown that changing the power-law dependence
of repulsion in a soft-sphere mixture leaves fragility
invariant~\cite{demichele04}, while altering both repulsive and
attractive parts of the Lennard-Jones potential can change
the fragility of a supercooled mixture~\cite{bordat04}. 
The role of intermolecular interactions should not, however, be
overemphasized. 
Statistical mechanics theories of the glass-transition often use
correlation functions or other coarse-grained information on the liquid properties as input, rather than
the bare interaction potential. For instance, the most celebrated mode
coupling theory~\cite{goetze99} emphasizes the role of pair
correlations, as described by the static structure factors. Providing
correlations between fragility and 
properties such as local order, which may be used to characterize 
mesoscopic domains in supercooled liquids, would also be more helpful for
understanding experimental trends.  

In this paper, we investigate the connection between fragility and
local order by performing Molecular Dynamics simulations of different
supercooled Lennard-Jones mixtures. We consider a set of equimolar,
additive mixtures with varying size ratio, together with some prototypical
mixtures (Sec.~\ref{sec:model}). Different from most numerical
simulations on Lennard-Jones mixtures, we cool these systems at constant
pressure (Sec.~\ref{sec:sim}). The isobaric fragilities obtained from
relaxation times and diffusion coefficient show a systematic 
variation in additive mixtures upon varying
size ratio~(Sec.~\ref{sec:fragility}). Different 
fragility indexes are also found for two
models~\cite{ka1,wahnstrom} that have often 
been employed in computational studies on the
glass-transition. We rationalize these
findings by analyzing the role of locally preferred structures, as 
identified by means of a Voronoi construction~(Sec.~\ref{sec:structure}). We
show that super-Arrhenius 
behavior of dynamical properties can be ascribed to a rapid growth,
upon supercooling, of slow domains possessing distinct microstructural
features. As a keynote, we find that the fragility of binary
mixtures is correlated to the thermal rate of growth of such domains: the
more fragile the mixture, the more rapid the increase of the
fraction of particles forming locally preferred structures. 

\section{Model binary mixtures}\label{sec:model} 

We have performed extensive Molecular Dynamics simulations for 
13 binary Lennard-Jones mixtures. All models are composed 
of $N=500$ classical particles enclosed in a cubic box with periodic
boundary conditions. Particles interact via the Lennard-Jones potential
\beq \label{eqn:lj}
u_{\alpha\beta}(r) = 4 \epsilon_{\alpha\beta} \left[ {\left( 
    \frac{\sigma_{\alpha\beta}}{r} \right)}^{12} -
  {\left( \frac{\sigma_{\alpha\beta}}{r} \right)}^6 \right]
\eeq
where $\alpha,\beta=1,2$ are indexes of species. In our convention, 
particles of species 2 have a smaller diameter than those of species 1
($\sigma_{22}<\sigma_{11}$), and we fix $\sigma_{11}=1.0$ for all systems. 
Reduced units will be used in the
following, assuming $\sigma_{11}$, $\epsilon_{11}$ and
$\sqrt{m_1\sigma_{11}^2/\epsilon_{11}}$ as units of
distance, energy and time respectively. 
We have employed the cutoff scheme of Stoddard and
Ford~\cite{stoddard73}, which ensures continuity up to the first derivative 
of $u_{\alpha\beta}(r)$ at the cutoff radius $r_c=2.5$. 

The interaction parameters of the mixtures considered in this work are
shown in Table~\ref{tab:sys}. 
These models are characterized by a varying
degree of non-additivity in the composition rule for the cross
interaction and by different values of size ratio
$\lambda=\sigma_{22}/\sigma_{11}$. Deviations from 
Lorentz-Berthelot composition rules can be quantified using 
\beq\label{eqn:lbrule}
\eta=\frac{\sigma_{12}}{(\sigma_{11}+\sigma_{22})/2} \qquad
\xi=\frac{\epsilon_{12}}{\sqrt{\epsilon_{11}\epsilon_{22}}}
\eeq
We will now proceed to a brief presentation of the relevant features of
these force fields.  

\subsection*{BMLJ -- Binary Mixture of Lennard-Jones particles}
This is the classic mixture of Kob and Andersen~\cite{ka1}, which has
been used extensively as a
model supercooled liquid. It is characterized by a significant
asymmetry in the interaction parameters, both in
the interaction diameters and in the energy scales of
the two species. Furthermore, cross interactions are strongly
non-additive ($\eta=0.85$, $\eta=2.1$). The number concentration of
large particles is fixed at $x_1=0.8$, as in the original work.

\subsection*{\NiY -- Lennard-Jones model for \NiY alloy}
The parametrization of this mixture has been introduced by
Della Valle~\etal~\cite{dellavalle94} to provide a realistic description of the
structural features of binary amorphous alloys of Ni and Y atoms. The
cross-interaction diameter is non-additive ($\eta=0.91$), as in the
case of BMLJ, but a single energy scale is present. The masses of the
two chemical species are equal. The number concentration $x_1=0.67$
allows deep supercooling of the mixture.

\subsection*{WAHN -- Additive mixture of Wahnstr\"om}
Introduced by Wahnstr\"om~\cite{wahnstrom}, this equimolar mixture is
actually the Lennard-Jones version of the supercooled soft-sphere
model used in the early simulations of Bernu~\etal~\cite{bernu87}. It has
been employed several times in the literature as a model
glass-former. The interaction parameters are additive and
characterized by a moderate size asymmetry ($\lambda=0.837$). Note
that the mass ratio is $m_2/m_1=0.5$. 

\subsection*{AMLJ-$\lambda$ -- Additive Mixture of Lennard-Jones particles}
This is a set of equimolar additive mixtures. The masses of the two
species are equal $m_1=m_2=1.0$ and the size ratio $\lambda$ is allowed
to vary, keeping $\sigma_{11}$ fixed at 1.0. Recently, Lennard-Jones
clusters of this type has been investigated~\cite{doye05}. Ten
different values of $\lambda$ have been used in the range
$0.60\leq \lambda\leq 1.0$. Note that AMLJ-0.837 would be the same as
the WAHN mixture, if not for the the different mass ratio.

\begin{table}[!tbp]
\caption{\label{tab:sys}
  Parameters of Lennard-Jones potentials for binary
  mixtures. Also shown are the masses $m_1$ and $m_2$ of the two species and the
  concentration $x_1$ of particles of species 1. In the case of additive
  mixtures AMLJ-$\lambda$, the following values of size ratio $\lambda$
  have been used: 0.60, 0.64, 0.70, 0.73, 0.76, 0.82, 0.88,
  0.92, 0.96, 1.00. 
}
\begin{ruledtabular}
\begin{tabular}{ccccc}
                & BMLJ & \NiY     & WAHN  & AMLJ-$\lambda$ \\    
\cline{2-5} 
$\sigma_{11}$   & 1.0  & 1.0     & 1.0   & 1.0 \\ 
$\sigma_{12}$   & 0.8  & 0.7727  & 0.916 & $(\lambda+1)/2$ \\ 
$\sigma_{22}$   & 0.88 & 0.6957  & 0.833 & $\lambda$ \\ 
$\epsilon_{11}$ & 1.0  & 1.0     & 1.0   & 1.0\\ 
$\epsilon_{12}$ & 1.5  & 1.0     & 1.0   & 1.0 \\ 
$\epsilon_{22}$ & 0.5  & 1.0     & 1.0   & 1.0 \\ 
$m_1$           & 1.0  & 1.0     & 1.0   & 1.0\\ 
$m_2$           & 1.0  & 1.0     & 0.5   & 1.0 \\
$x_1$           & 0.8  & 0.67    & 0.5   & 0.5 \\
\end{tabular}


\end{ruledtabular}
\end{table}

\section{Quenching protocols and simulation details}\label{sec:sim}

\begin{figure*}
\includegraphics*[width=0.96\textwidth]{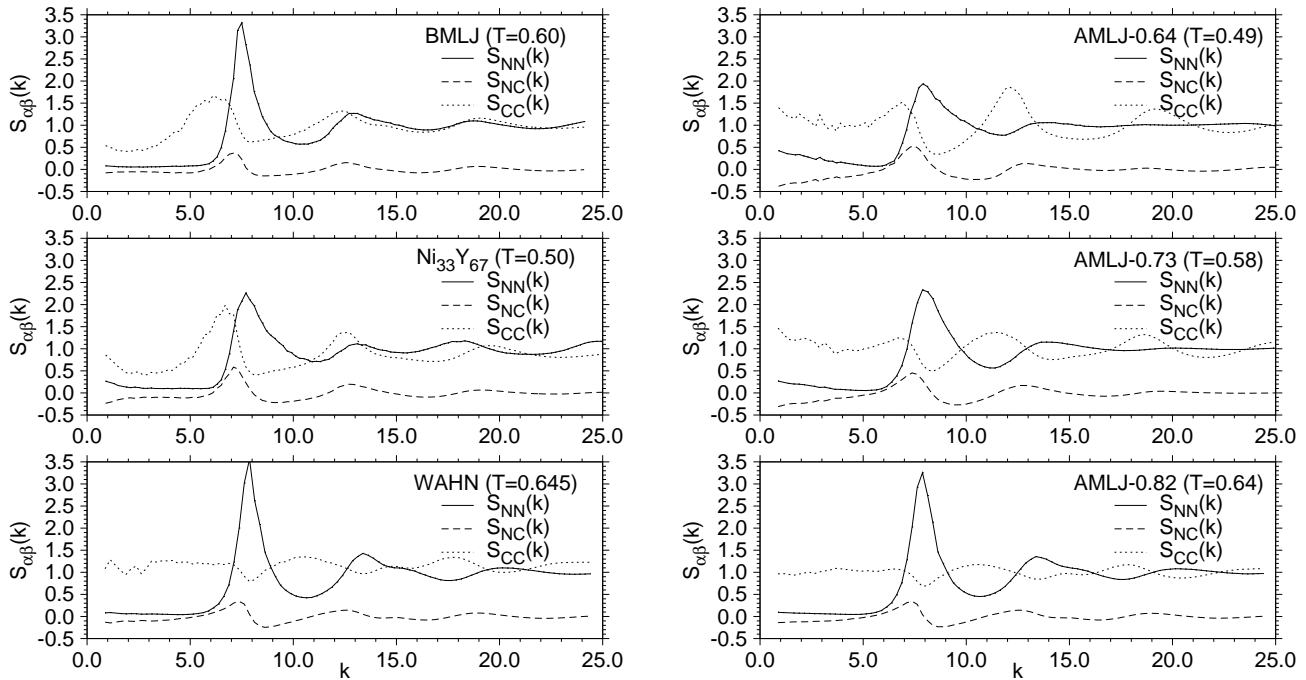}
\caption{\label{fig:sknc}
  Bathia-Thornton structure factors in the deeply supercooled
  regime. Number-number (solid lines), number-concentration (dashed
  lines) and concentration-concentration (dotted lines) structure
  factors are shown. 
  The concentration-concentration structure factor
  has been normalized to one by plotting $S_{CC}(k)/(x_1 x_2)$. 
  Data are shown at the lowest equilibrated temperature for each given system. 
  Note that all structure factors are finite in the limit
  $k\rightarrow 0$ and that the first sharp peak of $S_{NN}(k)$
  around $k_0\approx 8$ is roughly system independent.
} 
\end{figure*}

\begin{figure}
\includegraphics*[width=0.46\textwidth]{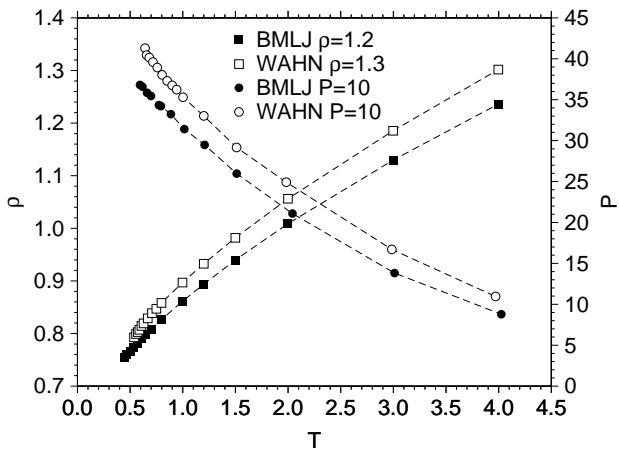}
\caption{\label{fig:presrho1}
  Temperature dependence of density $\rho(T)$ along isobaric
  quenches at $P=10$ (left axis, circles) and of pressure $P(T)$
  in isochoric quenches (right axis, squares). Data are shown
  for BMLJ (filled symbols) and WAHN (open symbols). Isochoric
  quenches were  performed at $\rho=1.2$ for BMLJ and $\rho=1.3$ for WAHN.} 
\end{figure}

Numerical simulations of model supercooled liquids are usually
performed at constant density. Recently, some attention has
been drawn to density effects~\cite{hernandezrojas03,tarjus04} and to
the role of pressure~\cite{mukherjee02,middleton03}. Nonetheless,
numerical studies of supercooled Lennard-Jones mixtures along isobaric quenches have seldom been
performed. In our case, following cooling paths at a common,
constant pressure provides a simple mean to perform a
homogeneous comparison of different mixtures. Moreover, this approach
is closer to the one usually employed in experimental
conditions, where most comparisons of supercooled liquids are
performed at constant (atmospheric) pressure. Whereas some minor differences
between isochoric and isobaric quenches can be observed, and will
be highlighted in the following, we remark that the main conclusions
of this work are independent of the choice of the quenching protocol.   

Isobaric quenches were performed by coupling the system to both
Berendsen thermostat and barostat during the equilibration phase. 
In most cases, we employed the standard Velocity-Verlet algorithm
for production runs. In order to achieve better control on temperature
in the deeply supercooled regime, we also performed a few
production runs
using the Nos\'e-Poincar\'e thermostat~\cite{bembenek,nose01}. 
For the system size and temperatures in consideration, simulations in different
ensembles provide consistent results, as far as average dynamical and
thermodynamical properties are concerned. 
Isochoric quenches were performed in a similar way,
using the Berendsen thermostat only during  equilibration.
The timestep $\delta t$ of both Velocity-Verlet and Nos\'e-Poincar\'e
integrators was varied from 0.002 at high temperature to
0.006 in the supercooled regime. The time constant for the Berendsen
thermostat~\cite{book:at} was $t_T=\delta t/0.1$, while the coupling
constant the for Berendsen barostat~\cite{book:at} was $10^3$ in reduced units. 
The inertia parameter of the Nos\'e-Poincar\'e thermostat was set to
$Q=5$. Thanks to the symplectic nature of the integrators and to the smooth
cutoff employed, no significant drift in the relevant conserved
Hamiltonian is observed in either kind of simulation used for
production runs, even for very long runs (up to $2.5\times 10^7$ steps
at the lowest equilibrated temperatures).

Beside checking the stability of mean potential
energy and pressure during the production phase, we adjusted the total
simulation times in order to achieve similar values 
of total root mean square displacement for all state points and
for all mixtures.  
Typical values of total root mean square displacement for large
particles are always larger than $4\sigma_{11}$.
To check the reliability of our results, we also tried both faster
and slower cooling rates, and performed some thermal histories by
reheating deeply supercooled samples. 
In the case of AMLJ-$\lambda$ mixtures, we found
that they could be safely supercooled for values of size ratio $\lambda$
between 0.60 and 0.84. In the case $\lambda=0.88$, in fact, some
early signs of crystallization were found in our sample. 
A part from this case, no sign of phase separation or crystallization
was detected looking at the time evolution of thermodynamic and structural properties.
The Bathia-Thornton~\cite{bathia70}  number-number, number-concentration and concentration-concentration
structure factors for a selection of mixtures are shown in
Fig.~\ref{fig:sknc} at the lowest equilibrated temperatures.

In the following, we will mostly consider isobaric quenches performed at a
reduced pressure $P=10$. Reference data for BMLJ are available along
this isobar~\cite{mukherjee02}, so that we could check the reliability of both
dynamical and thermodynamical properties obtained in our
simulations. In the case of BMLJ, we performed a few isobaric
quenches over a wider range of pressure ($P=5,10,20,50$) to
investigate the pressure dependence of isobaric fragility. Isochoric
quenches have been 
performed for BMLJ and WAHN 
mixtures, fixing the density at $\rho=1.2$ and $\rho=1.3$
respectively. These values of $\rho$ are equal to the ones used
in the original papers~\cite{ka1,wahnstrom}. In
Fig.~\ref{fig:presrho1} we show the temperature dependence
of density (at constant pressure) and pressure (at constant density)
for BMLJ and WAHN. Note that the range of pressure investigated for
BMLJ ($5\leq P\leq 50$) is consistent with the variation of pressure
for this system along the isochore $\rho=1.2$. 

\section{Fragility}\label{sec:fragility}

The fragility of a glass-former quantifies
how rapid the change of dynamical properties is upon
supercooling. We will focus on dynamical
quantities that can be computed with good statistical accuracy
in numerical simulations, namely relaxation times for the decay of 
density fluctuations and diffusion coefficients. 

\begin{table*}[htbp]
\caption{\label{tab:fragility}
  Fitted parameters for relaxation times $\tau(T)$ of large particles
  according to the generalized Vogel-Fulcher-Tammann given by
  Eq.~\eqref{eqn:arrvft},
  and for total diffusion coefficient $D(T)$ according to
  Eq.~\eqref{eqn:vftdiff}. The reference temperature 
  $T_r$ and the onset temperature $T_{onset}$ are described in the
  text.}  
\begin{ruledtabular}
\begin{tabular}{llllllllll}
    &    &                &     & \multicolumn{4}{c}{Relaxation times}
                                & \multicolumn{2}{c}{Diffusion coefficient} \\
\cline{5-8}
\cline{9-10}
    & P   & $T_{onset}$ & $T_r$ & $\tau_{\infty}$ & $E_{\infty}$ & $T_0$ & $K$ & $T_0$ & $K$ \\
\hline
      BMLJ &     5.0 & 0.75 & 0.464 &  0.0931(3) &    1.99(1) &   0.392(3) &    0.43(2) &   0.361(6) &    0.37(2) \\ 
      BMLJ &    10.0 & 0.95 & 0.574 &  0.0815(5) &    2.61(1) &   0.479(2) &    0.40(1) &   0.457(3) &    0.41(1) \\ 
      BMLJ &    20.0 & 1.20 & 0.765 &   0.067(1) &    3.71(9) &    0.63(1) &    0.40(5) &    0.60(1) &    0.42(3) \\ 
      BMLJ &    50.0 & 1.80 & 1.248 &  0.0481(9) &    6.60(7) &    1.03(1) &    0.38(3) &    1.01(1) &    0.46(2) \\ 
      WAHN &    10.0 & 0.90 & 0.623 &  0.0825(4) &    2.33(1) &   0.573(4) &    0.94(6) &   0.523(6) &    0.70(4) \\ 
      WAHN &    20.0 & 1.20 & 0.825 &  0.0670(6) &    3.38(3) &   0.752(5) &    0.84(5) &   0.697(9) &    0.68(4) \\ 
      \NiY &    10.0 & 0.90 & 0.489 &  0.0777(7) &    2.53(2) &   0.391(3) &    0.31(1) &   0.379(4) &    0.37(1) \\ 
 AMLJ-0.60 &    10.0 & 0.85 & 0.451 &   0.076(1) &    2.43(3) &   0.341(5) &    0.24(1) &   0.319(4) &    0.29(1) \\ 
 AMLJ-0.64 &    10.0 & 0.90 & 0.474 & 0.07691(3) &   2.444(1) &   0.381(4) &    0.32(1) &   0.356(4) &    0.36(1) \\ 
 AMLJ-0.70 &    10.0 & 0.90 & 0.514 &  0.0811(1) &   2.359(4) &   0.440(5) &    0.48(3) &   0.417(7) &    0.51(4) \\ 
 AMLJ-0.73 &    10.0 & 0.90 & 0.560 &  0.0785(7) &    2.48(2) &   0.502(4) &    0.71(4) &   0.466(6) &    0.62(4) \\ 
 AMLJ-0.76 &    10.0 & 0.90 & 0.601 &  0.0790(9) &    2.49(3) &   0.554(4) &    1.01(9) &   0.519(7) &    0.84(7) \\ 
 AMLJ-0.82 &    10.0 & 0.95 & 0.636 &  0.0803(5) &    2.53(1) &   0.591(6) &     1.1(1) &    0.53(1) &    0.76(7) \\ 
\hline
    & $\rho$  & $T_{onset}$ & $T_r$ & $\tau_{\infty}$ & $E_{\infty}$ & $T_0$ & $K$ & $T_0$ & $K$ \\
\hline
      BMLJ &     1.2 & 1.00 & 0.422 &   0.110(2) &    2.69(2) &   0.331(3) &    0.31(1) &   0.328(2) &   0.426(9) \\ 
      WAHN &     1.3 & 1.05 & 0.522 &   0.097(1) &    2.73(2) &   0.447(5) &    0.50(4) &   0.426(2) &    0.64(1) \\ 
\end{tabular}

\end{ruledtabular}
\end{table*}

For the definition of relaxation times we will 
consider the self part of the intermediate scattering function
\beq\label{eqn:fskt}
F_s^{\alpha}(k,t) = \frac{1}{N_{\alpha}}\sum_{i=1}^{N_{\alpha}} 
                    \lav \exp 
		    \left\{i \kve \cdot [\rve_i(t+t_0)-\rve_i(t_0)]\right\}
		    \rav
\eeq
where $\alpha=1,2$ is an index of species and 
$\lav\,\rav$ indicates an average over time origins $t_0$. Relaxation
times $\tau_{\alpha}$ for species $\alpha$ are defined by the condition
$F_s^{\alpha}(k^*,\tau_{\alpha})=1/e$, where $k^*$ corresponds to the
position of the first peak in the number-number structure factor
(see Fig.~\ref{fig:sknc}). The value of $k^*$ is roughly system- and temperature
independent for the mixtures in consideration, and close to $k\approx 8$. 
In the following, we will focus on the temperature
dependence of $\tau \equiv \tau_1$, but similar trends are
observed when considering the small particles. 

The difficulty of providing an unbiased, global description of the
temperature dependence of transport coefficients and relaxation times
by fitting the experimental data has been particularly stressed by
Kivelson \etal~\cite{kivelson96}. Care must be taken when the
definition of fragility itself relies on a specific functional form,
or when the latter is used for extrapolations outside the accessible
range of temperature.
We thus seek functional forms that are reliable over a large range of
temperature and require the range
for fitting to be well-specified and physically motivated.
For describing the temperature dependence of relaxation times, we
start with the well-known Vogel-Fulcher-Tammann (VFT) law and write it
in the form~\cite{sastry98} 
\beq \label{eqn:vft}
\tau(T)=\tau_{\infty}\exp\left[\frac{1}{K(T/T_0-1)}\right]
\eeq   
The material-dependent parameter $K$ quantifies the fragility of
the glass-former under consideration. The larger is $K$, the steeper is the increase of 
$\tau(T)$ upon supercooling. 
Equation~\eqref{eqn:vft} provides a fairly good description of relaxation
times in the deeply supercooled regime, but is inaccurate at high
temperature~\cite{kivelson96}. In the normal liquid regime, in fact,
relaxation times have a mild temperature dependence, which is well
described by the Arrhenius law. The existence of a crossover between 
these two regimes around some temperature $T_{onset}$, accompanied by  
several qualitative changes in the properties of the liquid, is well
established in the literature~\footnote{The crossover temperature 
$T_{onset}$ can be identified by the onset of non-exponential,
two-step relaxation in $F_s^{\alpha}(k^*,t)$~\cite{ka2}. In constant
density simulations, 
it is also signaled by the sharp decrease of average
potential energy of local minima~\cite{sastry98}. For further
interpretation regarding $T_{onset}$, see Ref.~\onlinecite{brumer04}.}.   
It seems thus sensible to use the following global functional form 
\beq\label{eqn:arrvft}
\tau(T) = 
\left\{ 
\begin{array}{ll}
  \tau_{\infty}     \exp\left[E_{\infty}/T\right] & T>T^*   \\
  \tau_{\infty}^{'} \exp\left[\dfrac{1}{K(T/T_0-1)}\right] & T<T^* \\
\end{array}
\right.
\eeq
where
\beq
\tau_{\infty}^{'}=\tau_{\infty}
\exp\left[E_{\infty}/T^{*} - \dfrac{1}{K(T^{*}/T_0-1)}\right]
\eeq
as a generalized VFT law. This functional form is continuous at the
crossover temperature $T^*$ and provides a fragility index $K$ with
analogous physical meaning to that in Eq.~\eqref{eqn:vft}.
To fit our simulation data to Eq.~\eqref{eqn:vft}, we proceed
similarly to Kivelson~\etal~\cite{kivelson96}. First  
we fit the relaxation times to the Arrhenius law
$\tau_{\infty}\exp(E_{\infty}/T)$ in the range $T>T_{onset}$ and then
we use $\tau_{\infty}$ and $E_{\infty}$ as fixed parameters in a
global fit to Eq.~\eqref{eqn:arrvft}.  In this way, a
good overall fit is obtained for relaxation times. Note that $T^*$ is
considered as a fitting parameter, but the main conclusions of this
section will not be altered by fixing $T^*=T_{onset}$.

\begin{figure}
\includegraphics*[width=0.46\textwidth]{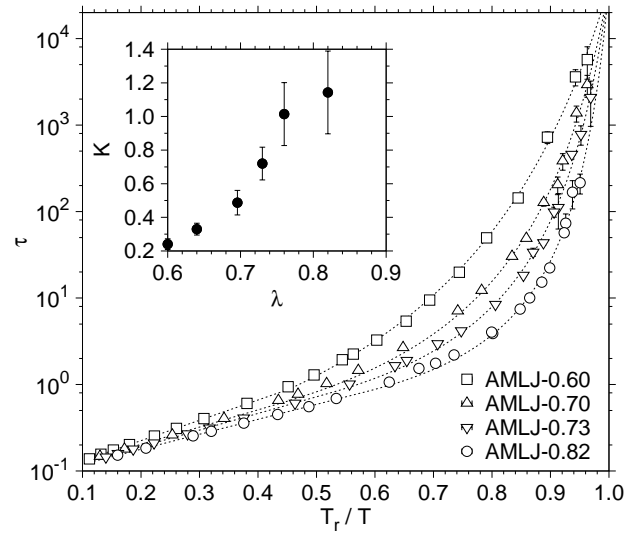}
\caption{\label{fig:arrtau3}
  Angell plot of relaxation times of large particles $\tau$
  for a selection of AMLJ-$\lambda$
  mixtures. Results are shown for 
  $\lambda=0.60,0.70,0.73,0.82$ along the isobar $P=10$. The 
  reference temperature $T_r$ is 
  described in the text. The inset shows the isobaric fragility index
  $K$ obtained from generalized VFT equation (see Eq.~\eqref{eqn:arrvft})
  against size ratio~$\lambda$.}   
\end{figure}

\begin{figure}
\includegraphics*[width=0.46\textwidth]{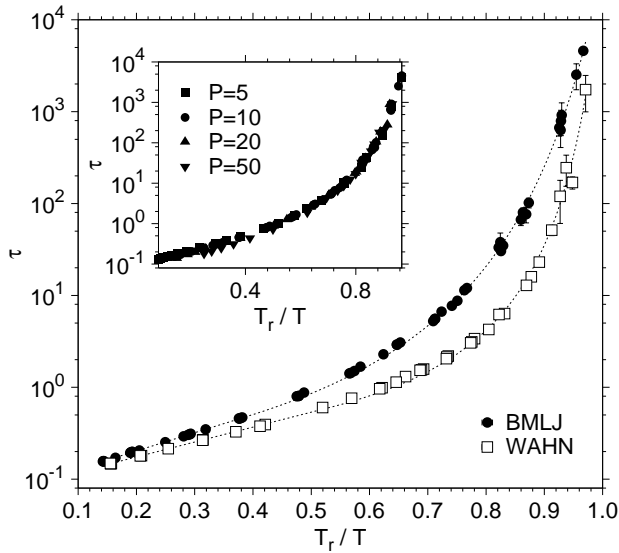}
\caption{\label{fig:arrtau4}
  Angell plot of relaxation times of large particles $\tau$ for
  BMLJ (black circles) and WAHN mixture (white circles) along isobaric
  quenches at $P=10$. The inset shows results at $P=5,10,20,50$ for
  BMLJ.}   
\end{figure}

The Angell plots of relaxation times in
Fig.~\ref{fig:arrtau3} and Fig.~\ref{fig:arrtau4} 
constitute the starting point of our discussion. For their construction  
we have used a reference temperature $T_r$, akin to the glass
transition temperature $T_g$, at which $\tau(T_r)=4 \times 10^4$. Such
a value of $\tau(T_r)$ is close to the one used by
Bordat~\etal~\cite{bordat04} in a study of modified Lennard-Jones
mixtures. For each mixture, the value of the
reference temperature $T_r$, which we have extrapolated using
Eq.~\eqref{eqn:arrvft}, is only slightly below the lowest
equilibrated temperature.

In Fig.~\ref{fig:arrtau3} we consider the set of additive, equimolar
mixtures AMLJ-$\lambda$ along
isobaric quenches at $P=10$. The size ratio
$\lambda$ is varied in the range $0.60\leq\lambda \leq 0.82$.
A strong, systematic variation is apparent upon varying $\lambda$: the
mixture becomes more fragile as $\lambda$ increases, 
i.e. as the size asymmetry between the two species is reduced.  
Recently, a similar influence of size ratio on fragility has been
observed in modified BMLJ mixtures~\cite{sun06}.
The trend of variation of fragility is confirmed by
our fitting procedure, whose outcome is summarized in
Table~\ref{tab:fragility}. The dependence of the 
isobaric fragility index $K$ on $\lambda$, shown in the inset of
Fig.~\ref{fig:arrtau3}, also suggests the existence of a saturation of
fragility around $\lambda=0.80$. This feature will be further
discussed in Sec.~\ref{sec:structure}, in connection with icosahedral
ordering. 

Figure~\ref{fig:arrtau4} shows the temperature dependence of relaxation
times for the two well-studied glass-formers BMLJ and WAHN, both
cooled isobarically at $P=10$. These mixtures have been used extensively
for numerical investigations of the glass-transition, but a direct
comparison has never appeared in the literature. 
We find that WAHN is appreciably more fragile than BMLJ,
independent of quenching protocols and system size~\footnote{We
observed a similar difference in fragility on both smaller
($N=108$) and larger samples ($N=6912$) cooled at constant density.}. 
The enhanced fragility of WAHN 
is not surprising since this mixture is, a part from a different mass
ratio, an AMLJ-$\lambda$ mixture with $\lambda=0.837$. We found that such
difference in mass ratio is irrelevant to the dynamical
properties in consideration. Thus, WAHN can be considered as the
end-point of a series of supercooled mixtures with increasingly large fragility.  

Does isobaric fragility itself depend on pressure? This
question has received much attention in the last years, in particular
within the experimental
community~\cite{paluch00,casalini05a,casalini05b,reiser06,niss06}. A tentative answer  
can be given for BMLJ, for which we performed isobaric quenches
in the range $5\leq P \leq 50$. By looking at the inset of
Fig.~\ref{fig:arrtau4}, we can see that relaxation times obtained
along different isobars collapse on a master curve by scaling $T$ with
the corresponding $T_r$. Numerical values of isobaric
fragility $K$, obtained from Eq.~\eqref{eqn:arrvft} along different
isobars, are also very close to each other (see
Table~\ref{tab:fragility}). Thus, our results indicate that the  
pressure dependence of isobaric fragility of Lennard-Jones mixtures
might be mild or even negligible for moderate variations of $P$. For a
given system, we also find that the isochoric fragility is slightly
smaller than the corresponding isobaric fragility, in agreement with
experimental observations~\cite{casalini05b}. 

\begin{figure}
\includegraphics*[width=0.46\textwidth]{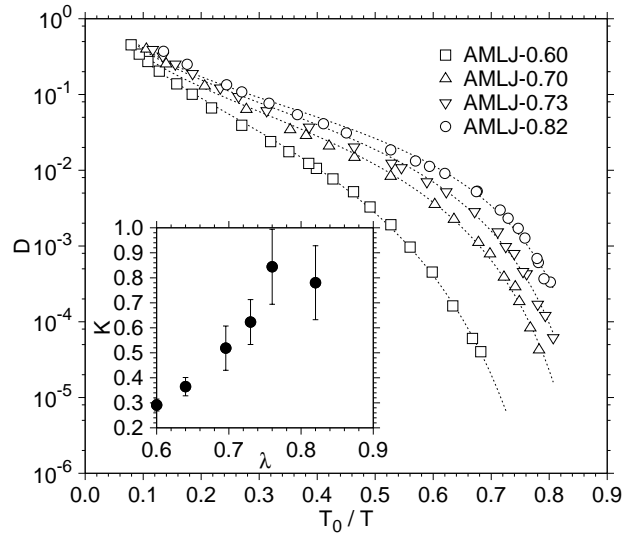}
\caption{\label{fig:arrdiff1}
  Angell plot of total diffusion coefficient $D$
  for a selection of AMLJ-$\lambda$ mixtures.
  Results are shown for 
  $\lambda=0.60,0.70,0.73,0.82$ along the isobar $P=10$. The reference
  temperature $T_0$ is obtained from fit to
  Eq.~\eqref{eqn:vftdiff}. The inset shows the 
  isobaric fragility index $K$ obtained from Eq.~\eqref{eqn:vftdiff}
  against size ratio $\lambda$.}
\end{figure}

\begin{figure}
\includegraphics*[width=0.46\textwidth]{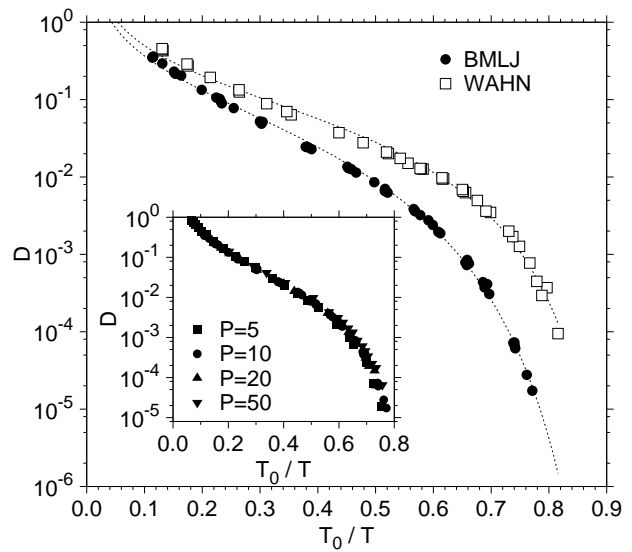}
\caption{\label{fig:arrdiff2}
  Angell plot of total diffusion coefficient $D$ for
  BMLJ (black circles) and WAHN mixture (white circles) along isobaric
  quenches at $P=10$. The inset shows results at $P=5,10,20,50$ for
  BMLJ.}   
\end{figure}

As another dynamical indicator we consider the the total diffusion
coefficient $D=x_1 D_1 + x_2 D_2$, given by the usual Einstein
relation. The temperature dependence of $D$ can be described quite
satisfactorily over the entire temperature range by a modified VFT
law~\cite{bordat03} 
\beq \label{eqn:vftdiff}
D(T)=D_0 T \exp\left[-\frac{1}{K(T/T_0-1)}\right]
\eeq   
Again, the parameter $K$ in Eq.~\eqref{eqn:vftdiff} provides
a measure of fragility. Angell plots for the diffusion coefficient
are shown in Figs.~\ref{fig:arrdiff1} and~\ref{fig:arrdiff2}, and results of fits to
Eq.~\eqref{eqn:vftdiff} are collected in
Table~\ref{tab:fragility}. Regarding the variation of fragility in
our models, results obtained for the diffusion coefficient confirm
the analysis based on relaxation times. Note that, by considering the
temperature dependence of both $\tau\equiv\tau_1$ and $D$, we account
for a possible dependence of fragility on chemical species and $k$-vector. 
Thus, the following trends appear to be rather robust:
(i)~The increase of fragility with size ratio $\lambda$ in 
AMLJ-$\lambda$ mixtures. (ii)~The less fragile behavior of BMLJ
compared to WAHN. (iii)~The independence of isobaric fragility on
pressure in BMLJ.

\section{Locally preferred structures}\label{sec:structure}

The growth of long-lived, slow domains,
characterized by well-defined structural features, has been discussed recently 
 in connection to super-Arrhenius behavior of dynamical
properties of supercooled liquids~\cite{shintani06,tarjus05,dzugutov02,doye03}. In this
section, we will follow a similar approach by analyzing the properties
of local order in the Lennard-Jones mixtures introduced in
Sec.~\ref{sec:model}. We will highlight the existence of different,
well-defined local geometries in these systems and show how their
properties are related to the variations of fragility.  

\begin{table}[!h]
\caption{\label{tab:voronoi}
  Most frequent Voronoi polyhedra around small particles. The
  percentage is computed with respect to the number of small
  particles in the system. Also shown is the average number of
  neighbors of species 
  1 ($n_1$) and 2 ($n_2$). Results refer to local minima along the isobar $P=10$ and are shown
  for $T=2.0$ and slightly above the reference temperature $T_r$,
  i.e. for the lowest equilibrated temperature.} 
\begin{ruledtabular}
\begin{tabular}{lrlccrlcc}
       & \multicolumn{4}{c}{$T=2.00$} &  \multicolumn{4}{c}{$T\approx T_r$}\\
\cline{2-5}
\cline{6-9}
       & \% & Signature & $n_1$ & $n_2$ & \% & Signature  & $n_1$ & $n_2$\\
\hline
AMLJ-0.64 & 12.0 &(0,2,8,1)& 7 & 4 	 & 13.9 & (0,2,8,1)& 7 & 4 \\
 & 7.3 &(0,2,8,2)& 6 & 6 	 & 9.3 & (0,2,8,2)& 7 & 5 \\
 & 7.1 &(0,2,8)& 7 & 3 	 & 8.3 & (0,2,8)& 7 & 3 \\
 & 5.8 &(0,3,6,3)& 7 & 5 	 & 6.7 & (0,3,6,3)& 7 & 5 \\
\hline
AMLJ-0.70 & 11.8 &(0,2,8,1)& 7 & 4 	 & 12.1 & (0,0,12)& 5 & 7 \\
 & 10.2 &(0,2,8,2)& 7 & 5 	 & 11.8 & (0,2,8,2)& 7 & 5 \\
 & 5.9 &(0,3,6,3)& 7 & 5 	 & 11.6 & (0,2,8,1)& 7 & 4 \\
 & 5.0 &(0,3,6,4)& 7 & 6 	 & 6.7 & (0,3,6,4)& 7 & 6 \\
\hline
AMLJ-0.82 & 14.3 &(0,0,12)& 6 & 6 	 & 29.1 & (0,0,12)& 6 & 6 \\
 & 10.9 &(0,2,8,2)& 7 & 5 	 & 10.8 & (0,2,8,2)& 7 & 5 \\
 & 7.3 &(0,3,6,4)& 7 & 6 	 & 8.1 & (0,1,10,2)& 7 & 6 \\
 & 6.7 &(0,1,10,2)& 7 & 6 	 & 6.9 & (0,3,6,4)& 8 & 5 \\
\hline
WAHN & 14.6 &(0,0,12)& 6 & 6 	 & 31.4 & (0,0,12)& 6 & 6 \\
 & 10.7 &(0,2,8,2)& 7 & 5 	 & 10.1 & (0,2,8,2)& 7 & 5 \\
 & 7.5 &(0,3,6,4)& 7 & 6 	 & 8.6 & (0,1,10,2)& 7 & 6 \\
 & 7.1 &(0,1,10,2)& 7 & 6 	 & 7.1 & (0,3,6,4)& 8 & 5 \\
\hline
BMLJ & 13.7 &(0,2,8)& 9 & 1 	 & 18.6 & (0,2,8)& 10 & 0 \\
 & 7.4 &(1,2,5,2)& 9 & 1 	 & 7.3 & (1,2,5,3)& 10 & 1 \\
 & 7.3 &(0,3,6)& 9 & 0 	 & 6.1 & (1,2,5,2)& 9 & 1 \\
 & 5.4 &(0,3,6,1)& 9 & 1 	 & 6.0 & (0,3,6)& 9 & 0 \\
\hline
NiY & 8.5 &(0,3,6)& 7 & 2 	 & 14.0 & (0,3,6)& 8 & 1 \\
 & 6.6 &(0,3,6,1)& 8 & 2 	 & 9.3 & (0,3,6,1)& 8 & 2 \\
 & 5.2 &(0,4,4,3)& 8 & 3 	 & 8.6 & (0,2,8)& 9 & 1 \\
 & 5.1 &(0,2,8)& 8 & 2 	 & 6.1 & (1,2,5,2)& 9 & 1 \\
\end{tabular}

\end{ruledtabular}
\end{table}

For this kind of analysis we have employed a Voronoi
construction~\cite{voronoi08}. Each particle in the system is the center of a Voronoi 
polyhedron, which is constructed by intersection of planes orthogonal to
all segments connecting the central particle to the other ones. 
Planes are drawn at a fraction $f_{\alpha\beta}$ of these
segments, where $\alpha$ is the species of the central
particle and $\beta$ is the species of the other particle.
We have used the recipe $f_{\alpha\beta} =
\sigma_{\alpha\alpha}/(\sigma_{\alpha\alpha} + 
\sigma_{\beta\beta})$~\cite{dellavalle94}, but the main conclusions
of this section will not be altered when considering the more intuitive choice,
$f_{\alpha\beta}=1/2$. The sequence $(n_3,n_4,\dots)$, where
$n_k$ is the number of faces of the polyhedron having $k$ vertices,
provides a detailed description of the local geometry around a given
particle. All zero values behind the maximum number of vertices of a
polyhedron are ignored. 
We have applied this kind of analysis to both instantaneous
configurations, sampled along Molecular Dynamics trajectories, and to
local minima of the potential energy, obtained by conjugate-gradients
minimizations. Between 200 and 2000 independent configurations have
been analyzed for each state point. We found that it is easier to
characterize local order in the Lennard-Jones mixtures considered in
this work by taking ``the point of view'' of small particles. Well-defined geometries
appear, in fact, most frequently around small particles, while no 
recognizable local order is apparent around large particles. In the
following, we will thus concentrate our attention on the properties of
Voronoi polyhedra centered around small particles.

Such a Voronoi construction will provide an indication of
what are the locally preferred structures~\cite{mossa03,tarjus05} of our model supercooled
liquids. In the rest, we will use an effective, purely geometric 
definition of locally preferred structures, as the ones corresponding
to the most frequent polyhedra found in our Voronoi construction.
Determining unambiguously the \bem origin \eem of the preference for a
given local structure in the bulk will require a significant
additional effort, since, in general, such preference will depend in a
non-trivial way on the environment surrounding a given local structure
and may be triggered by factors other than energetic stability. For
instance, packing effects 
can play an important role in stabilizing a local structure. Moreover,
in the case of binary mixtures, compositional freedom further
increases the complexity of such analysis. Taking into account the
effect of the liquid environment around a local structure at a
mean-field level, and identifying the appropriate ``local free energy''
to be minimized, remains an open issue of current
research~\cite{mossa03,mossa06}. Here, we tackle these issues by using a
purely geometric definition of locally preferred 
structures, which has the further advantage of being available even
when energetic criteria will fail, e.g. for hard-spheres. 

\begin{figure}
\includegraphics*[width=0.46\textwidth]{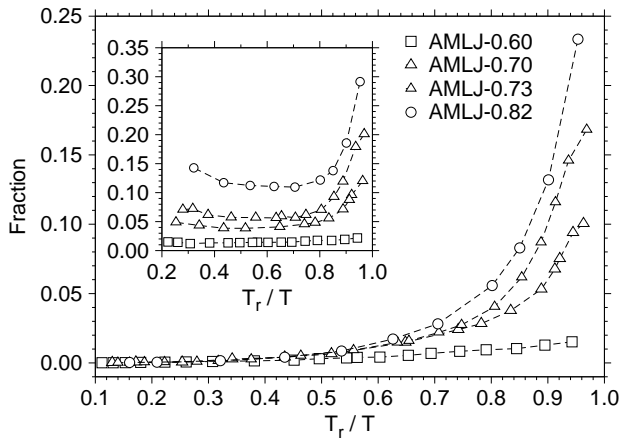}
\caption{\label{fig:voronoifraction1} 
  Temperature dependence of the fraction of small
  particles at the center of (0,0,12)-polyhedra in AMLJ-$\lambda$ for
  selected values of $\lambda$. Results are shown for instantaneous
  configurations (main plot) and local minima (inset) along isobaric
  quenches at $P=10$.} 
\end{figure}

Let us first focus on equimolar, additive mixtures AMLJ-$\lambda$. One of the
most relevant structural features of these mixtures is the existence
of a varying degree of icosahedral ordering.
Icosahedral coordination has a sharp signature in the Voronoi
construction, being associated to (0,0,12)-polyhedra, i.e. 12
pentagonal faces.
The temperature dependence of the fraction of icosahedra for different
AMLJ-$\lambda$ mixtures, shown in Fig.~\ref{fig:voronoifraction1},
displays a striking correlation with the variation of fragility. In
fact, the increase of icosahedral ordering upon supercooling is more 
rapid and more pronounced as $\lambda$ increases, i.e. as the fragility
of the mixture increases~\footnote{The fraction of icosahedra in local
minima also shows, at intermediate temperatures, a shallow minimum,
which is more pronounced the more fragile is the mixture, and
appears to be a distinct feature of constant pressure simulations.}. 
To our knowledge, this is the first time that such a relationship
is established in supercooled binary mixtures. Previous numerical
studies have focused, in fact, on the connection between icosahedral
ordering and super-Arrhenius behavior in monoatomic
liquids~\cite{dzugutov02,doye03}. The theoretical interpretation of our
results is by no means trivial. On the one hand, the variation of
fragility with icosahedral ordering may be understood within the
frustration-limited domains theory~\cite{tarjus05} in terms of a more
rapid stabilization, upon supercooling, of locally 
preferred structures ---in the present case, icosahedra---. We will
further discuss this point below. On the other
hand, the trend we find in our simulations and shown in
Fig.~\ref{fig:voronoifraction1} appears to be at variance with the
results of a recent phenomenological model~\cite{tanaka05b,tanaka05c}.
Further clarifications on the relevance of this theoretical approach
to our simulated systems are required.  

\begin{figure}
\includegraphics*[width=0.46\textwidth]{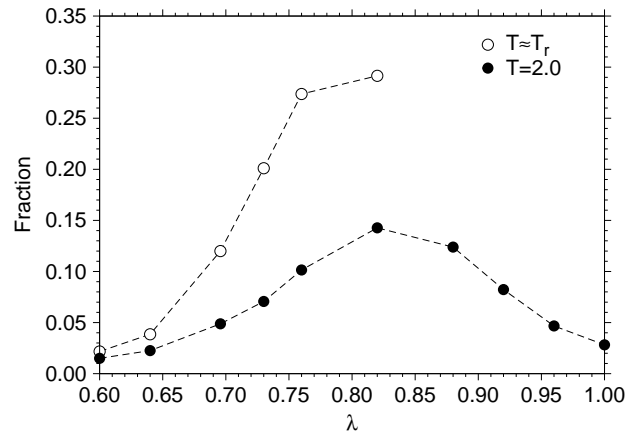}
\caption{Variation of icosahedral ordering with size ratio in additive
  mixtures AMLJ-$\lambda$. The fraction of small particles at the
  center of (0,0,12)-polyhedra in local minima is shown as a function of
  size ratio $\lambda$, at $T=2.0$ (black circles) and at the lowest
  equilibrated temperatures (open circles).}   
\label{fig:voronoi1}
\end{figure}

To have a better feeling of how icosahedral ordering is triggered by
size ratio, we show, in Fig.~\ref{fig:voronoi1}, the fraction of
(0,0,12)-polyhedra in local minima as a function of $\lambda$. Results are shown along
the isotherm $T=2.0$ and for $T \approx T_r$, i.e. at the lowest 
temperatures that could be accessed in equilibrium condition. 
Depending on temperature, a different range of $\lambda$ is
considered. In the deeply supercooled regime ($T\approx T_r$) 
only mixtures with $0.60\leq\lambda\leq 0.84$ could be equilibrated
(see Sec.~\ref{sec:sim}). 
For variation of $\lambda$ in this range, icosahedral ordering
increases with size ratio, in a way which strongly resembles the
increase of fragility index $K$ with $\lambda$, reported in
Sec.~\ref{sec:fragility}. At high temperature
($T=2.0$) the full range $0.60\leq\lambda\leq 1.00$ can be
accessed and our data reveal the existence of a maximum of
icosahedral ordering around $\lambda\approx 0.82$. This feature might
provide a simple explanation to the existence of a saturation of 
fragility around $\lambda=0.80$ reported in Sec.~\ref{sec:fragility}. 
Interestingly, the results obtained from local minima at high temperature
show that the variation of icosahedral ordering with size ratio, which
is apparent in the deeply supercooled regime,
is already encoded in the liquid inherent structure~\cite{sw82}. 

\begin{figure}
\includegraphics*[width=0.42\textwidth]{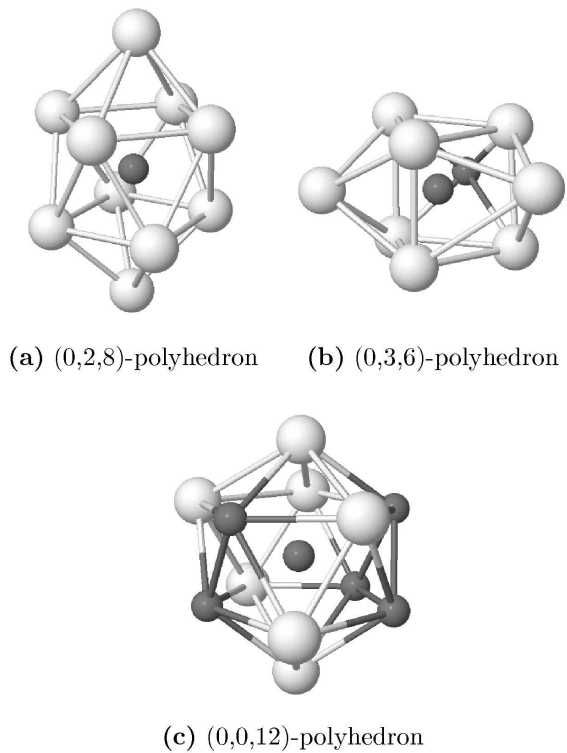}
\caption{Examples of locally preferred structures found in local
  minima of supercooled Lennard-Jones mixtures. Small and large particles are shown as dark
  and pale spheres respectively. (a) (0,2,8)-polyhedron (twisted bicapped square
  prism) in BMLJ. This is the most frequent chemical coordination, incidentally
  one or two small particles can form the cap. (b) (0,3,6)-polyhedron
  (capped trigonal prism) in \NiY. In this case, one of the caps is often formed
  by a small particle. (c) (0,0,12)-polyhedron (icosahedron) in
  WAHN. On average, the coordination around the central particle is
  equimolar.} 
\label{fig:voronoisingle}
\end{figure}

Such a pattern of variation of icosahedral 
ordering with size ratio is strikingly similar to that observed in 
models of bidisperse Cu glasses~\cite{lee03} and in the realistic models
of metallic glasses developed by Hausleitner and
Hafner~\cite{hausleitner92a,hausleitner92b}. This suggests that 
the increase of icosahedral ordering with size ratio, and its
consequent correlation with fragility, might be a general feature of
binary systems with additive, or nearly additive, spherical
interactions~\footnote{In the models of Hausleitner and
Hafner, the interaction parameters become non-additive at large size
asymmetry, favoring prismatic geometries. In the case of additive
AMLJ-$\lambda$ mixtures, instead, no sharp structural characterization
is apparent at large size asymmetry.}.
Furthermore, the onset of crystallization for $\lambda\agt
0.88$ could be simply be explained, in this kind of systems, by the
decrease of icosahedral coordination and by a larger occurrence of
(0,3,6,4)-polyhedra and (0,4,4,6)-polyhedra, which are typical of FCC
crystals~\cite{tanemura77}. We found, in fact, that 
the fraction of FCC-related polyhedra in the normal liquid regime
increases as $\lambda\rightarrow 1$. A similar behavior has been
explicitly demonstrated for the bidisperse Cu model by means of a
Honeycutt-Andersen construction.    

We can now generalize the connection between fragility and locally
preferred structures by analyzing systems possessing favored
geometries different from icosahedra. Such opportunity is provided by
non-additive mixtures, such as BMLJ and \NiY. 
In the case of \NiY, in fact, icosahedral ordering has been shown to
be strongly frustrated~\cite{dellavalle94}. Non-additivity of the
interaction potential favors the formation of trigonal
prismatic structures~\cite{dellavalle94},
similar to those found in the crystalline phases of
NiY alloys. This is confirmed by our analysis, which shows (see
Table~\ref{tab:voronoi}) that the
most frequent Voronoi polyhedron around small particles in \NiY  
is the (0,3,6)-polyhedron, corresponding to capped
trigonal prismatic structures. 
In the case of BMLJ, 
we find that the (0,2,8)-polyhedron has the largest occurrence around
small particles both at high and low temperature (see
Table~\ref{tab:voronoi}). This polyhedron corresponds to twisted
bicapped square prisms, mostly formed by neighboring large particles. 
The preference for twisted prismatic structures in BMLJ has also been
highlighted in studies on the coordination polyhedra in the 
liquid~\cite{fernandez04}, and on the stability of isolated
clusters~\cite{doye07}.  
As a working hypothesis, we identify the geometries associated to
(0,3,6)-polyhedra and (0,2,8)-polyhedra as the locally preferred
structures of \NiY and BMLJ, respectively. These two non-additive
mixtures can be effectively contrasted to WAHN, which displays
a strong icosahedral ordering upon supercooling. 

\begin{figure}
\includegraphics*[width=0.46\textwidth]{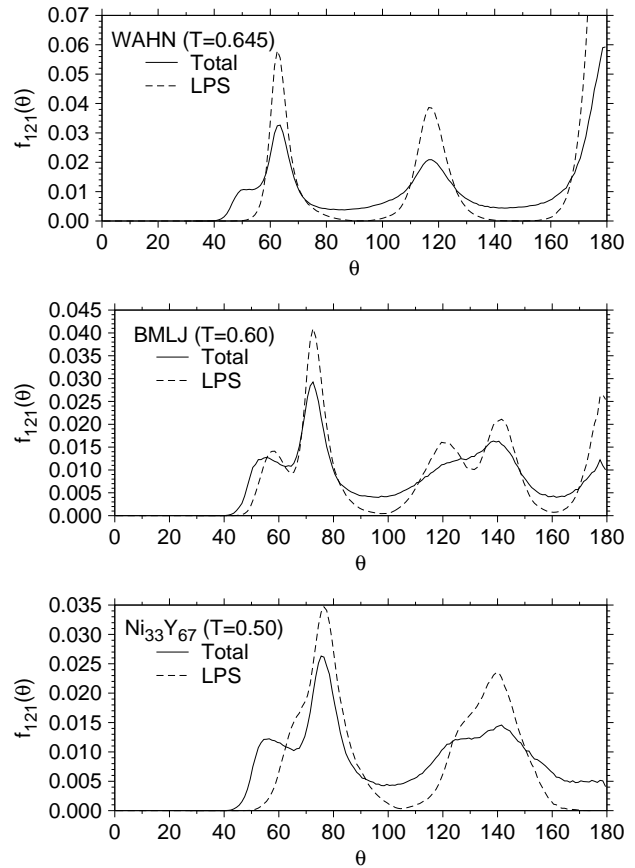}
\caption{
  Bond-angle distributions around small particles for WAHN (upper
  plot), BMLJ (middle plot) and \NiY (lower plot). The bond-angle distribution
  $f_{121}(\theta)$ is shown as solid line. Also shown is the
  bond-angle distribution $f_{121}(\theta)$ restricted to small
  particles which are at the center of the locally preferred
  structure of the system, as given by
  Fig.~\ref{fig:voronoisingle}. Data refer to the lowest equilibrated  
  temperature of each given system. The sharp peaks in the 
  $f_{121}(\theta)$ distributions filtered for locally preferred
  structures reflect the ideal angles of the corresponding geometry.}
\label{fig:angle}
\end{figure}

In Fig.~\ref{fig:voronoisingle} we show three highly symmetric
configurations corresponding to locally preferred structures, found in
local minima of WAHN, BMLJ and \NiY. Notice that the structures shown in the figure 
are among the most symmetric in their own class of Voronoi polyhedra.
Support to our definition of locally preferred structures is
given by the analysis of angular distributions.
To this aim, we compute the bond-angle distribution functions
$f_{\alpha\beta\gamma}(\theta)$ around particles of species $\alpha$, where
$\beta$ and $\gamma$ are the species of the neighboring particles. 
Particles sharing a face in the Voronoi construction are considered as
neighbors. In Fig.~\ref{fig:angle} we focus on the bond-angle distribution
$f_{121}(\theta)$  for central small particles and 
neighboring large particles in WAHN, BMLJ and \NiY. We compare the total
$f_{121}(\theta)$ to the one restricted to small particles being the
center of the typical Voronoi polyhedron of the mixture. A strong
resemblance is observed between the average environment around small
particles in the bulk and the local geometries of Voronoi
polyhedra corresponding to our putative locally preferred structures. 

\begin{figure}
\includegraphics*[width=0.46\textwidth]{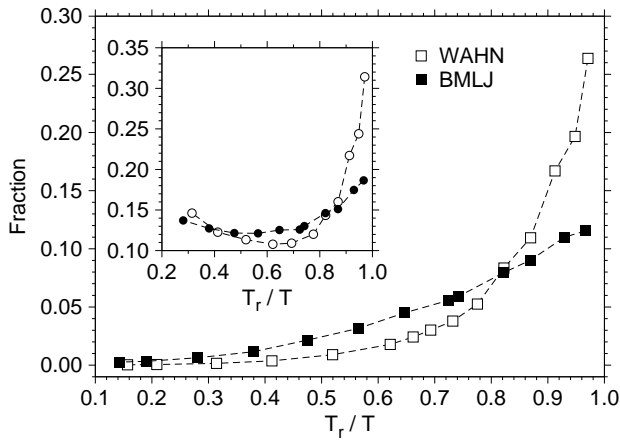}
\caption{\label{fig:voronoifraction2} 
  Temperature dependence of the fraction of small
  particles at the center of selected Voronoi polyhedra in
  instantaneous configurations (main plot) and local minima (inset).
  The fraction of (0,2,8)-polyhedra in BMLJ (white symbols) and
  (0,0,12)-polyhedra in WAHN mixture (black symbols) are
  shown along isobaric quenches at $P=10$. Data for \NiY are close to
  those for BMLJ, but are not shown for clarity.}
\end{figure}

\begin{figure}
\includegraphics*[width=0.40\textwidth]{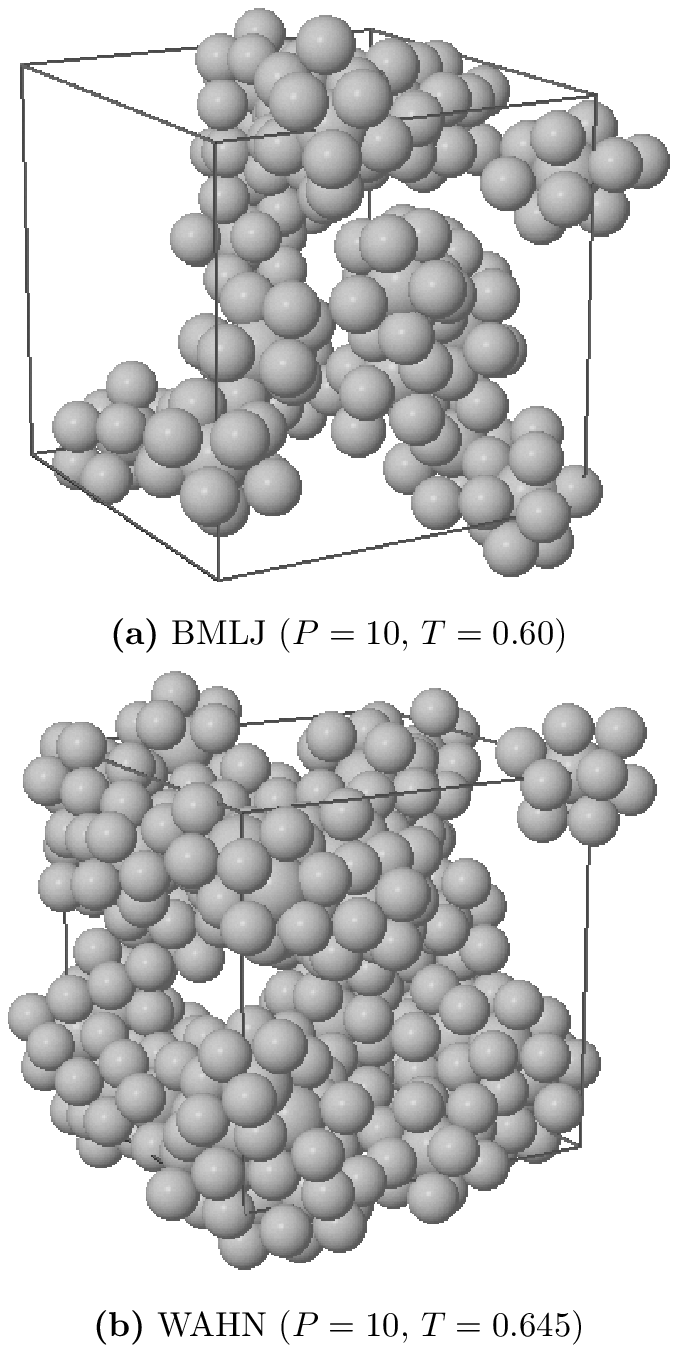}
\caption{\label{fig:voronoifull}
  Domains formed by locally preferred structures in local minima at
  the lowest equilibrated temperature at $P=10$ (WAHN: $T=0.645$. BMLJ: $T=0.60$).
  Particles forming (a) (0,2,8)-polyhedra in BMLJ and (b)
  (0,0,12)-polyhedra in WAHN are shown as spheres of the same radius,
  irrespectively of chemical species.}   
\end{figure}

As it can be seen from Fig.~\ref{fig:voronoifraction2}, the thermal rate of
growth of the fraction of particles forming locally preferred structures 
is again correlated to the fragility of the model. In fact, the
fraction of icosahedra in WAHN increases rapidly by decreasing
temperature, whereas the growth of prismatic structures, typical of
BMLJ and \NiY, is rather mild. 
The most frequent Voronoi polyhedra of all these mixtures are not homogeneously spread in
the system. They tend to form growing domains as the temperature
decreases.  This feature is exemplified by the two snapshots in
Fig.~\ref{fig:voronoifull}, where we show the typical extension of
domains formed by locally preferred structures in local minima, for
deeply supercooled BMLJ and WAHN mixtures.
Similar extended domains formed by interlocking icosahedra have
been found in the supercooled regime of Dzugutov liquids~\cite{dzugutov02,doye03}.

\begin{figure*}
\includegraphics*[width=0.98\textwidth]{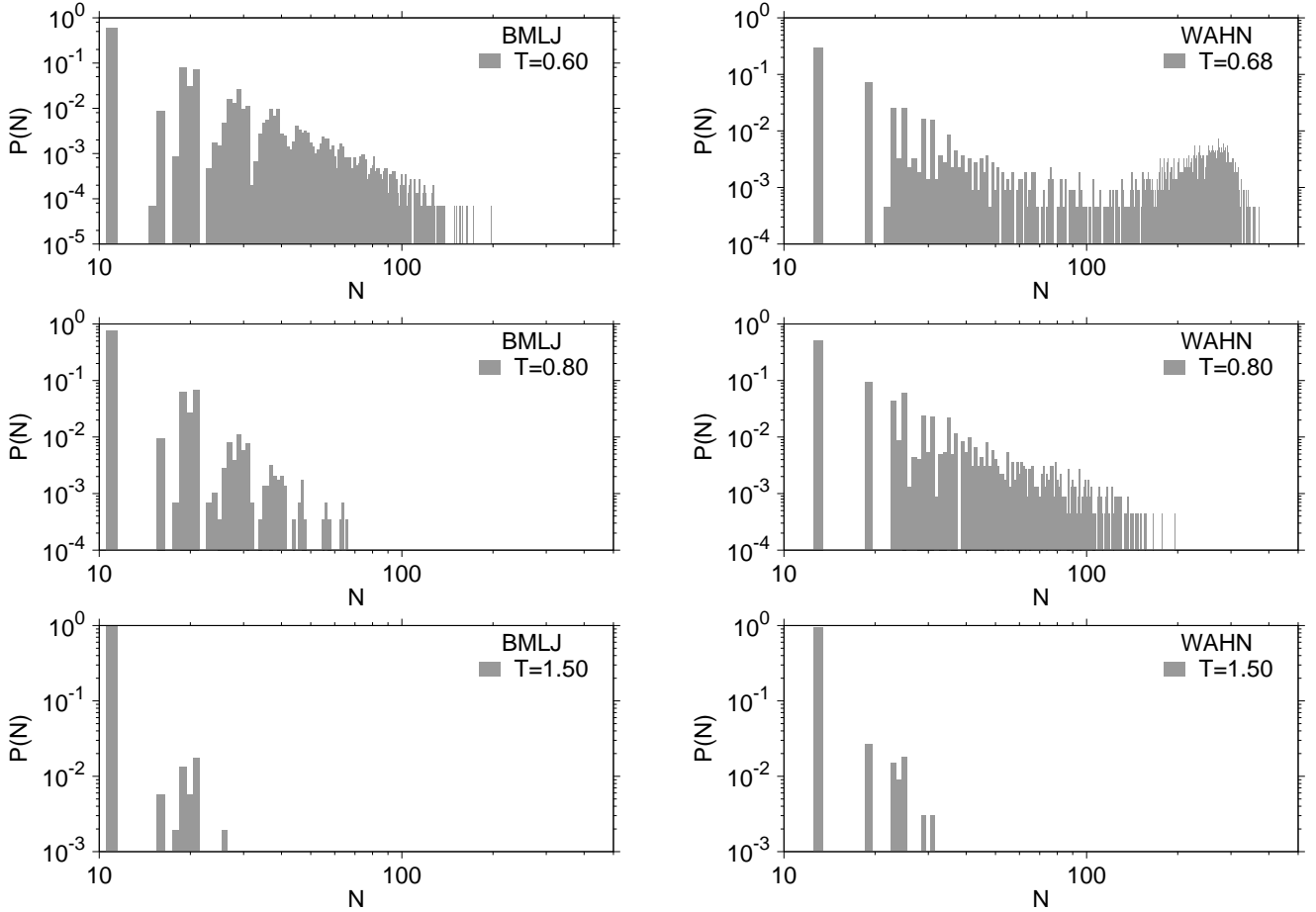}
\caption{\label{fig:voronoicluster}
  Distribution $P(N)$ of the size $N$ of domains formed by locally
  preferred structures in BMLJ (left plots) and WAHN (right plots) at
  different $T$. Results refer to isobaric quenches at $P=10$.
}   
\end{figure*}

To assess the statistical relevance of the presence of such domains,
we analyze the distribution $P(N)$ of clusters composed by $N$
neighboring particles forming locally preferred structures. 
Our identification of domains is as follows. For each given configuration,
we partition the particles into three classes: (i)
$c$-particles, which are the center of a locally 
preferred structure; (ii) $n$-particles, which are neighbors to some
other $c$-particle, but are not themselves centers of a locally
preferred structure; (iii) $o$-particles, which are neither
$c$-particles nor $n$-particles, i.e. they are outside the
domains formed by locally preferred structures. Particles sharing a
face in the Voronoi construction are then considered as
neighbors. Using the partitioning scheme above, we identify domains as
clusters composed by neighboring $c$- and $n$- particles. 
The distribution $P(N)$ is shown in
Fig.~\ref{fig:voronoicluster} for instantaneous configurations of BMLJ
and WAHN at different state 
points. By decreasing temperature, a clear tendency of forming
larger clusters is observed in both systems. In the WAHN mixture we find
that, around $T_r$, there is almost always a large cluster formed by
icosahedra percolating in the simulation box, beside some smaller
ones. This feature is reflected in the bimodal distribution of $P(N)$ for WAHN at low
temperature. On the other hand, as expected from the analysis of
Fig.~\ref{fig:voronoifraction2}, the growth of domains in BMLJ
appears to be more limited. 

\begin{figure}
\includegraphics*[width=0.46\textwidth]{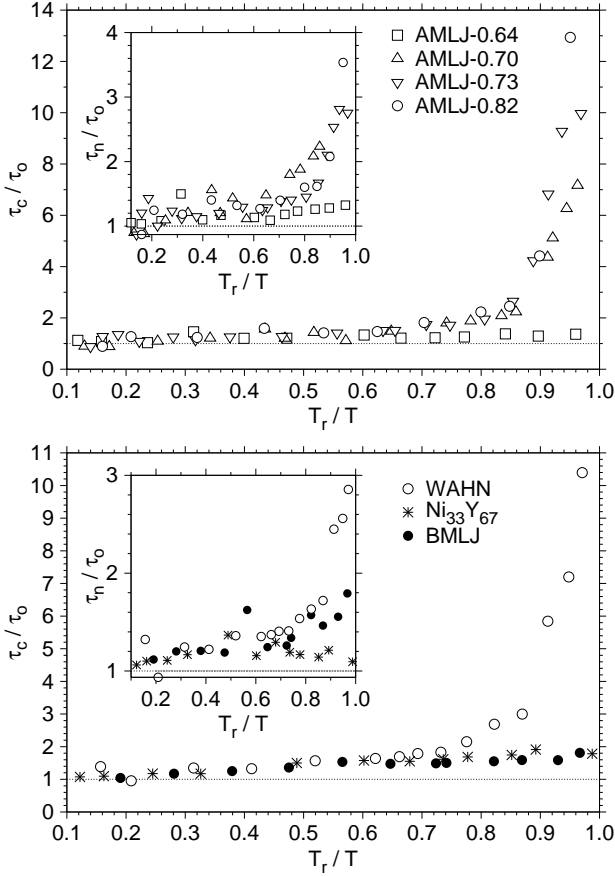}
\caption{\label{fig:voronoi_arrtau}
  Dynamical impact of locally preferred structures, as
  identified by the temperature dependence of the ratio
  $\tau_c/\tau_o$ (main plot) and $\tau_n/\tau_o$ (insets) at $P=10$. See text
  for definition of $\tau_c$, $\tau_n$, and $\tau_o$. Upper
  plot: AMLJ-$\lambda$ mixtures for $\lambda=0.60,0.70,0.73,0.82$. Lower
  plot: BMLJ (filled circles), WAHN (open circles) and \NiY
  (stars). The dotted line drawn at 1 indicates the high temperature
  limit.}    
\end{figure}

Such domains are expected to provide an efficient
mechanism of slowing down in supercooled liquids~\cite{dzugutov02}.
To address this issue, we evaluate the self intermediate scattering
function for small particles [see Eq.~\eqref{eqn:fskt}], according to their
role in the locally preferred structure at time $t=t_0$. 
At each time origin $t_0$, we partition the \bem small \eem particles
into $c$-, $n$-, and $o$-particles, as described above.
The correlation functions
$F^c_s(k,t)$, $F^n_s(k,t)$, and $F^o_s(k,t)$ are then obtained by
performing the average over time origins in Eq.~\eqref{eqn:fskt} using
only $c$-, $n$-, or $o$-particles respectively.
Relaxation times $\tau_c$, $\tau_n$, and $\tau_o$ are defined as in
Sec.~\ref{sec:fragility}. 
The ratio $\tau_c/\tau_o$ provides a simple measure of the slowness of
particles inside domains formed by a particular locally
preferred structure. 
The temperature dependence of $\tau_c/\tau_o$ is
shown in Fig.~\ref{fig:voronoi_arrtau} for various mixtures at $P=10$. The value of $\tau_c/\tau_o$
tends to 1 at high temperature in all systems and increases more
markedly by decreasing temperature, as the fragility of the system increases. Around $T_r$, we find that
relaxation times within icosahedral domains differ by roughly an order
of magnitude from those outside, whereas prismatic structures in
non-additive mixtures develop a more modest separation of
time scales. 

\begin{table}[!h]
\caption{\label{tab:voronoi_lifetime}
  Lifetime $\tau_p$ of most frequent Voronoi polyhedra around small particles.
  Results are obtained from local minima at the lowest equilibrated
  temperatures ($T\approx T_r$). Also shown is the ratio
  $\tau_p/\tau_2$, where $\tau_2$ is the relaxation time obtained from
  the condition $F_s^2(k^*,\tau)=1/e$.
} 
\begin{ruledtabular}
\begin{tabular}{llrr}
       & \multicolumn{3}{c}{$T\approx T_r$}\\
\cline{2-4}
       & Signature & $\tau_p$ & $\tau_p/\tau_2$ \\
\hline
WAHN & (0,0,12) & 2000 & 1.6 \\ 
   & (0,1,10,2) & 90 & 0.1\\ 
   & (0,2,8,2)  & 60 & 0.0\\ 
   & (0,3,6,4)  & 60 & 0.0\\ 
\hline
BMLJ & (0,2,8) & 800 & 0.4 \\ 
   & (0,3,6)   & 200 & 0.1\\ 
   & (1,2,5,3) & 70 & 0.0\\ 
   & (1,2,5,2) & 40 & 0.0\\ 
\hline
\NiY & (0,2,8) & 2500 & 0.5 \\ 
   & (0,3,6) & 1600 & 0.3\\ 
   & (0,3,6,1) & 1300 & 0.3\\ 
   & (1,2,5,2) & 90 & 0.0\\ 
\end{tabular}

\end{ruledtabular}
\end{table}

The dynamical impact of locally preferred structures is assisted, at
low temperature, by an increased lifetime of such slow domains. To address this point we
proceeded similarly to Donati~\etal~\cite{donati99}, introducing a
single-particle function $\nu_i(t)$ that equals 1 if 
particle $i$ is at the center of a given Voronoi polyhedron,
and 0 if not. Restricting our attention to small particles, we
computed the autocorrelation function~\cite{donati99} 
\beq
\sigma(t)=\sum_{i=1}^{N_2} \langle\nu_i(t)\nu_i(0)\rangle
         -\frac{n_p^2}{N_2}
\eeq
where $n_p=\sum_{i=1}^{N_2} \langle\nu_i(0)\nu_i(0)\rangle$ is the
average number of small particles at the center of a given
polyhedron. We estimated the lifetime $\tau_p$ of a
polyhedron from the condition $\sigma(\tau_p)=\sigma(0)/e$.
Independent of the polyhedron under consideration, the normalized
autocorrelation function $\sigma(t)/\sigma(0)$ falls rapidly to zero
in the normal liquid regime. As the temperature is lowered, polyhedra
corresponding to locally preferred structures become more long-lived
than the others, as expected. Around $T_r$, we find that
$\sigma(t)/\sigma(0)$ for locally preferred structures decays to zero within the
time scale given by the decay of $F_s^2(k^*,t)$. In
Table~\ref{tab:voronoi_lifetime}, we report the lifetimes $\tau_p$ of some
frequent polyhedra found at the lowest equilibrated temperatures for
WAHN, BMLJ, and \NiY. In WAHN and BMLJ, the lifetime of polyhedra
corresponding to locally preferred structures is around an order of magnitude
larger than those of other geometries. Interestingly, in the case of
\NiY, we find that some less frequent polyhedra, such as
(0,2,8)-polyhedra, have a lifetime comparable to that of our putative
locally preferred structure, suggesting the existence of competing
structures. We also find that icosahedra tend to 
have a longer lifetime, relative to the typical structural relaxation
times, than prismatic structures.

The relation between fragility and local order, which is apparent from
our simulation data, fits rather well into the scenario of the
frustration-limited domains theory~\cite{tarjus05}. According to this approach,
glass-formation 
arises from the competition of a tendency to form mesoscopic, stable
domains, characterized by locally preferred structures, and a mechanism
of frustration, which prevents these domains from tiling three
dimensional space. Fragility turns out to be proportional to the
energetic stability of such domains and inversely proportional 
to the strength of frustration. At present, the roles of
stability and frustration cannot be clearly
disentangled. Nevertheless, the following considerations, based on the present work, are possible and we
hope they could serve as guidelines for further theoretical modeling or
investigations: (i)~In the case of additive 
mixtures, within the explored range of size ratio, icosahedral ordering seems to be
the most prominent structural feature. 
Results obtained for isolated Lennard-Jones clusters~\cite{doye05}
suggest that the maximum of icosahedral ordering around $\lambda\approx
0.84$ might be related to an enhanced energetic stability of
equimolar icosahedra, i.e. icosahedra with the same number of large
and small neighbors. Formation of more stable icosahedral structures would, in
turn, explain the increase of fragility with size ratio. A more
detailed study of larger clusters in the bulk, forming extended regions
of icosahedral coordination~\cite{doye03,dzugutov02}, would probably be
required to further clarify this point. 
(ii)~It should be realized that frustration in
different systems may be of different origin. Icosahedral ordering, while being frustrated 
from tiling three-dimensional space for geometrical reasons, 
allows the growth of relatively
large domains, when compared to prismatic structures observed in
non-additive Lennard-Jones mixtures. 
In non-additive alloys different competing locally preferred structures and mismatch in stoichiometry
may further increase frustration.

\section{Conclusions}

In this work, we have provided a comparative study of different
supercooled Lennard-Jones mixtures by quenching the systems at constant
pressure. The models analyzed include the well-studied
mixtures BMLJ~\cite{ka1} and WAHN~\cite{wahnstrom}, a set of equimolar, additive 
mixtures with varying size ratio, and a model meant the
give a realistic structural description of the \NiY alloy~\cite{dellavalle94}. 
These systems display a varying degree of fragility, which has been
rationalized in terms of the properties of some relevant locally  
preferred structures.   

Local order has been characterized using a Voronoi construction. 
Employing an effective definition of the locally preferred
structure of a liquid, as the geometry corresponding
to the most frequent Voronoi polyhedra, we have shown that fragility
is related, in the mixtures considered in this work, to the rapid
growth with temperature of slow, stable domains 
characterized by the locally preferred structure of the
mixture, generalizing previous observations on monoatomic
bulk systems~\cite{dzugutov02,doye03}. Such a growth with temperature 
is more rapid, the more fragile the mixture. 

Analyzing the set of AMLJ-$\lambda$ mixtures, we found
that the size ratio $\lambda$ controls the formation of icosahedral
ordering in the bulk. Extended regions of icosahedral coordination are more
rapidly formed upon 
supercooling of additive mixtures with moderate size asymmetry. This,
in turn, leads to a more pronounced super-Arrhenius behavior. 
These results might also be representative of a
wider class of mixtures with additive, non-directional interactions, since
icosahedral ordering has been found to display a similar trend of
variation with size ratio in models of bidisperse Cu glasses~\cite{lee03}.  
On the other hand, non-additive mixtures, such as BMLJ and \NiY, favor
the formation of prismatic structures. The growth, upon supercooling, of these structures
is milder compared to that of icosahedra in additive
mixtures. Consequently, non-additive mixtures display a less fragile
behavior, which might be due to the presence of stronger
frustration mechanisms. 
From such discussion of our data, it is thus tempting to relate fragility
to the thermal rate of growth of locally preferred structures,
whatever their type may be. 

In the light of the frustration-limited domains
theory~\cite{tarjus05}, fragility should originate from the interplay
between stability of domains formed by locally favored geometries and
frustration. Our results indicate that further studies are
needed to assess the relative role of these two factors in
supercooled mixtures. 
Deeper investigations on the link
between fragility and local order, either by study of isolated
clusters~\cite{doye05} or by direct determination of locally preferred
structures in the bulk~\cite{mossa03,mossa06} 
or by confocal microscopy technique~\cite{conrad06},
will certainly be rewarding.    

\begin{acknowledgments}
The authors would like to thank R.~G. Della Valle for making available to  
us his efficient program for Voronoi analysis. 
Computational resources for the present work have been partly obtained
through a grant from  ``Iniziativa Trasversale di Calcolo Parallelo'' of
the Italian {\em CNR-Istituto Nazionale per la Fisica della Materia}
(CNR-INFM) and partly within the agreement between the University of
Trieste and the Consorzio Interuniversitario CINECA (Italy).
\end{acknowledgments}

\bibliographystyle{apsrev}
\bibliography{references}

\end{document}